# Robust Capacity Assessment of Distributed Generation in Unbalanced Distribution Networks Incorporating ANM Techniques

Xin Chen, Wenchuan Wu, *Senior Member*, *IEEE*, Boming Zhang, *Fellow*, *IEEE*

*Abstract*—To settle a large-scale integration of renewable distributed generations (DGs), it requires to assess the maximal DG hosting capacity of active distribution networks (ADNs). For fully exploiting the ability of ADNs to accommodate DG, this paper proposes a robust comprehensive DG capacity assessment method considering three-phase power flow modelling and active network management (ANM) techniques. The two-stage adjustable robust optimization is employed to tackle the uncertainties of load demands and DG outputs. With our method, system planners can obtain the maximum penetration level of DGs with their optimal sizing and sitting decisions. Meanwhile, the robust optimal ANM schemes can be generated for each operation time period, including network reconfiguration, on-load-tap-changers regulation, and reactive power compensation. In addition, a three-step optimization algorithm is proposed to enhance the accuracy of DG capacity assessment results. The optimality and robustness of our method are validated via numerical tests on an unbalanced IEEE 33-bus distribution system.

## Nomenclature

### A. Set and Notation

| | |
|---|---|
| $N_{all}$ | Set of all nodes. |
| $N_{nr}$ | Set of nodes that are not root nodes. |
| $N_{dg}$ | Set of nodes potentially connected to distributed generations. |
| $N_{con}$ | Set of nodes connected to continuously adjustable VAR compensators. |
| $N_{dis}$ | Set of nodes connected to discretely adjustable VAR compensators. |
| $N_d(i)$ | Set of downstream nodes connected to node $i$. |
| $N(i)$ | Set of all nodes directly connected to node $i$. |
| $B_{all}$ | Set of all branches. |
| $B_{tr}$ | Set of branches with transformers. |
| $B_{nt}$ | Set of branches without transformers. |
| $\Gamma$ | Set of all time periods selected for DG capacity assessment. |
| $M_0$ | A very big positive number. |
| $\varphi$ | Phase index indicating phase A, B or C. |

### B. Parameter

| | |
|---|---|
| $\hat{\eta}_{i,t}^{\varphi}$ | Predicted efficiency coefficient of DG outputs at node $i$ at time $t$. |
| $\theta_{i,\min}^{\varphi}, \theta_{i,\max}^{\varphi}$ | Lower, upper limits of power factor angles of DG units at node $i$. |
| $q_{i,\min}^{c,\varphi}, q_{i,\max}^{c,\varphi}$ | Lower, upper limits of reactive power of continuously adjustable VAR compensators at node $i$. |
| $\Delta q_i^d$ | Reactive power increment per step of discretely adjustable VAR compensators at node $i$. |
| $\hat{p}_{i,t}^{L,\varphi}, \hat{q}_{i,t}^{L,\varphi}$ | Predicted active, reactive load demands at node $i$ at time $t$. |
| $s_{ij,\max}^{\varphi}$ | Apparent power capacity of branch $ij$. |
| $r_{ij}^{\varphi}, x_{ij}^{\varphi}$ | Self-resistance, self-reactance of branch $ij$. |
| $u_{i,\min}^{\varphi}, u_{i,\max}^{\varphi}$ | Lower, upper limits of squared voltage magnitude at node $i$. |
| $n_b$ | Number of all nodes. |
| $n_{root}$ | Number of root nodes. |
| $B_t$ | Uncertainty budget of time $t$. |

### C. Variable

| | |
|---|---|
| $y_i^{\varphi}$ | Installation DG capacity at node $i$. |
| $w_{ij,t}$ | Binary status variable of branch $ij$ at time $t$, equaling 1 if connected, otherwise 0. |
| $\rho_{ij,t}$ | Virtual power flow of branch $ij$ from node $i$ to node $j$ at time $t$. |
| $p_{i,t}^{dg,\varphi}, q_{i,t}^{dg,\varphi}$ | Active, reactive DG power injection at node $i$ at time $t$. |
| $q_{i,t}^{c,\varphi}$ | Reactive power injection of continuously adjustable VAR compensators at node $i$ at time $t$. |
| $q_{i,t}^{d,\varphi}$ | Reactive power injection of discretely adjustable VAR compensators at node $i$ at time $t$. |
| $\chi_{i,t}^{\varphi}$ | Nonnegative integer variable representing step number of discretely adjustable VAR compensators at node $i$ at time $t$. |
| $p_{ij,t}^{\varphi}, q_{ij,t}^{\varphi}$ | Active, reactive power flow of branch $ij$ from node $i$ to node $j$ at time $t$. |
| $v_{i,t}^{\varphi}$ | Voltage magnitude of node $i$ at time $t$. |

Manuscript received xx, 2017. This work was supported in part by the National Key Research and Development Plan of China (Grant 2016YFB0900400), and the National Science Foundation of China (Grant 51477083). The authors are with the Department of Electrical Engineering, Tsinghua University, Beijing 100084, China (email: 1689598375@qq.com; wuwench@tsinghua.edu.cn( Corresponding Author)).

| | |
|---|---|
| $u_{i,t}^{\varphi}$ | Squared voltage magnitude of node $i$ at time $t$. |
| $\tilde{\eta}_{i,t}^{\varphi}$ | Actual efficiency coefficient of DG outputs at node $i$ at time $t$. |
| $\tilde{p}_{i,t}^{L,\varphi}, \tilde{q}_{i,t}^{L,\varphi}$ | Actual active, reactive load demands at node $i$ at time $t$. |
| $\tau_{ij,t}$ | Turn ratio of OLTC at branch $ij$ at time $t$. |

*D. Three-phase Vector and Matrix*

| | |
|---|---|
| $\dot{\boldsymbol{v}}_{i,t}$ | Complex voltage vector of node $i$ at time $t$. |
| $\boldsymbol{u}_{i,t}$ | Squared voltage magnitude vector of node $i$ at time $t$. |
| $\dot{\boldsymbol{s}}_{ij,t}$ | Complex power flow vector of branch $ij$ from node $i$ to node $j$ at time $t$. |
| $\dot{\boldsymbol{s}}_{i,t}^{D}$ | Net load vector of node $i$ at time $t$. |
| $\dot{\boldsymbol{z}}_{ij}$ | Impedance matrix of branch $ij$. |

I. INTRODUCTION

In recent years, renewable energy resources based distributed generations (DGs) have proliferated rapidly in active distribution networks (ADNs), especially wind power generators and photovoltaic panels. However, when reaching a high penetration level, the integration of DG causes many emerging technical issues due to their non-dispatchable and intermittent generation mechanism. One of the most serious problems is over voltage violation, which turns out to be the main reason that restricts the usage of DGs. To cope with the adverse effects of DG integration, it requires a hosting capacity assessment process ahead of the practical deployment. In this way, system planners can figure out how much DG capacity is allowed to connect to present distribution networks, and determine the optimal sizes and locations of DG units. Essentially, the capacity assessment issue is an optimization problem to maximize DG hosting capacity while satisfying the system operating constraints. Hitherto, many studies have been conducted to this subject, including analytical approaches [1]-[3], heuristic methods [4]-[6] and optimal power flow (OPF) algorithms [7]-[9].

Via rescheduling network topology and adjusting power injection, the active network management (ANM) techniques can change power flow conveniently and effectively; they are powerful tools to mitigate voltage violation and enhance the amount of DG hosting capacity. The main ANM schemes involve network reconfiguration, on-load-tap-changer (OLTC) regulation, use of reactive power (VAR) compensators, and DG power factor control. Reference [10] [11] discussed the benefits of reactive power compensation, energy curtailment, and OLTC control for maximizing DG penetration level. In [12], the strategies of static and dynamic network reconfiguration were investigated to enlarge DG installation capacity. In these researches, the turn ratio of discretely adjustable OLTC is formulated as a continuous variable for simplification. Since OLTC regulation considerably affects the holistic voltage profiles of ADNs, it is better to use an accurate transformer model when assessing the DG hosting capacity.

Furthermore, load demands and outputs of renewable DGs still cannot be predicted accurately owing to their inherent volatility, especially for the long term. Limited real-time measurements in present distribution networks further exacerbate the estimation errors of load demands and DG outputs. Therefore, obvious errors of generations and loads bring significant uncertainties to the DG capacity assessment process. However, the uncertainties are ignored in the aforementioned works, which may lead to erroneous capacity evaluation results and inappropriate DG installation decisions. In researches [13] [14], masses of stochastic scenarios were generated based on historical data to simulate the real situation, and decisions were made with the stochastic optimization (SO) method. Generally, scenario based SO models can maintain the same mathematical formulations as the deterministic models, which are simple and straight. However, it is intractable to obtain the true probability distributions of uncertain variables in reality, and masses of scenarios usually impose a heavy computational burden for the optimization. As a good alternative to address uncertainties, robust optimization (RO) does not require the detailed information about probability distributions. Worst-case oriented RO can guarantee the feasibility of decisions under uncertainties, while it tends to come up with more conservative strategies than SO. Reference [15] proposed a maximum hosting capacity evaluation model with robust operation of OLTC and static VAR compensators (SVCs), which is formulated in a single-stage robust optimization framework and takes no account of network reconfiguration.

In addition, the power flow in distribution networks is intrinsically unbalanced due to non-symmetrical conductor, untransposed lines, and unequal three-phase loads [16]. A large-scale integration of DGs also aggravates the power imbalance for their unbalanced power injection in phases. Hence, the single-phase power flow model cannot match ADNs anymore, and a three-phase formulation is required for the DG hosting capacity assessment issue.

In summary, ANM schemes, uncertainties of generations and loads, and three-phase power flow model are decisive in determining the maximal hosting capacity and installation strategies of DGs, any of which turns out to be indispensable. For instances, if the uncertainty risks were ignored, it might make overly radical DG installation decisions, and then frequent voltage violation would occur in practical operation. On the contrary, if ANM techniques were not taken into consideration, it might come up with overly pessimistic capacity evaluation results and would be uneconomical. To avert biased evaluations, in this paper, we propose a robust comprehensive DG capacity assessment methodology, which incorporates all these three crucial elements simultaneously.

The major contributions of this paper are threefold and summarized as follows.

1) With our previous work [17], we develop an efficient three-phase linear branch power flow (LBPF) model with exact modelling for OLTC of transformers, which is suitable for the DG capacity assessment issue. To further enhance the accuracy of OPF models, we present another three-phase LBPF model considering network losses, and a three-step optimization algorithm incorporating these two LBPF models is proposed to perform DG capacity assessment efficiently and accurately.

2) Based on the two-stage robust optimization framework, we establish a robust comprehensive DG capacity assessment model (RC-CAM). In the first stage of RC-CAM, we make the optimal DG installation decisions and generate the optimal ANM strategies for each time period to maximize the total DG hosting capacity. In the second stage, the worst-case scenarios that jeopardize DG integration are sought within a predefined uncertainty sets. Since decisions are made under the worst-case scenarios, the DG capacity evaluation results obtained via RC-CAM can be immune to the impacts of forecasting errors.

3) To the best of our knowledge, this is the first work to propose such a robust comprehensive method for ADN optimization, in which all ANM schemes are simultaneously simulated with three-phase power flow models. Actually, it is a general method that can be extended to many other ADN optimization problems, such as voltage control, economic dispatch etc.

Besides, the uncertainty budget technique [18] is utilized to control the holistic conservativeness of RC-CAM, providing a trade-off between robustness and conservativeness.

The remainder of this paper is organized as follows: In Section II, we present the mathematical formulation of RC-CAM. In Section III, the column-and-constraint generation algorithm is applied to solve this two-stage robust model. In Section IV, we propose a three-step optimization algorithm to enhance the accuracy of assessment. Numerical tests are discussed in Section IV, and conclusions are drawn in Section V.

## II. MATHEMATICAL MODEL BUILDING

In this section, a three-phase LBPF model for unbalanced distribution networks is introduced at first, which presents an exact linearization method for OLTC modelling. Then, we come up with a deterministic capacity assessment model to evaluate the maximum DG hosting capacity of distribution systems. This model is deemed to be a comprehensive model for integration of all ANM strategies, including network reconfiguration, OLTC adjustment, and reactive output control of DGs and VAR compensation. After that, we employ robust optimization to tackle the uncertainties of loads and DG outputs, and propose the two-stage robust DG capacity assessment model based on this deterministic one.

### A. Three-phase Linear Branch Power Flow Model

In this paper, we denote a complex variable by the point sign, i.e., $\dot{s}_{ij,t} = p_{ij,t} + jq_{ij,t}$, and a notation without this point sign denotes the corresponding norm, i.e., $v_{i,t} = \|\dot{v}_{i,t}\|$. The bold lower case letters represent three-phase vectors or matrixes, and the thin superscripted indices with $\varphi = \{A,B,C\}$ denote their corresponding phase components, e.g., $\dot{s}_{ij,t} = \begin{bmatrix} \dot{s}_{ij,t}^{A} & \dot{s}_{ij,t}^{B} & \dot{s}_{ij,t}^{C} \end{bmatrix}^{T}$. The superscripted asterisk on vectors or matrixes indicates their conjugates, i.e., $\dot{v}_i^*$. Besides, $\otimes$ and $\oslash$ denote the element-wise multiplication and division, respectively. In addition, the variables with the superscript $\circ$ indicate their corresponding optimal values.

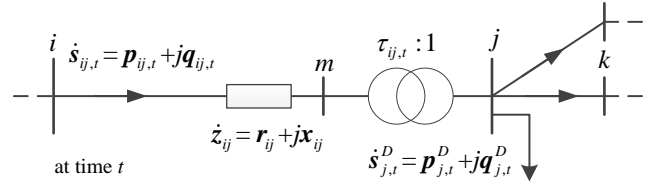

Figure 1. A branch of radial distribution networks with a transformer.

Consider a certain branch of radial distribution networks with a transformer, shown as branch $ij$ in Fig.1. For better analysis, we divide branch $ij$ into branch $im$ and branch $mj$. Branch $im$ has the same impedance as branch $ij$, while branch $mj$ only contains a tap-changer with turn ratio $\tau_{ij,t}$. At a certain time $t$, the three-phase power flow equations of branch $ij$ can be formulated as (1)-(3) exactly, whose derivation process is provided in Appendix A.

$$u_{i,t} - u_{m,t} = \dot{v}_{i,t} \otimes \left[ \dot{z}_{ij}^* \cdot \left( \dot{s}_{ij,t} \oslash \dot{v}_{i,t} \right) \right] + \left[ \dot{z}_{ij} \cdot \left( \dot{s}_{ij,t}^* \oslash \dot{v}_{i,t}^* \right) \right] \otimes \dot{v}_{i,t}^* \\ - \left[ \dot{z}_{ij} \cdot \left( \dot{s}_{ij,t}^* \oslash \dot{v}_{i,t}^* \right) \right] \otimes \left[ \dot{z}_{ij}^* \cdot \left( \dot{s}_{ij,t} \oslash \dot{v}_{i,t} \right) \right] \quad (1)$$

$$\dot{s}_{ij,t} - \left[ \dot{z}_{ij} \cdot \left( \dot{s}_{ij,t}^* \oslash \dot{v}_{i,t}^* \right) \right] \otimes \left( \dot{s}_{ij,t} \oslash \dot{v}_{i,t} \right) = \sum_{k \in N_d(j)} \dot{s}_{jk,t} + \dot{s}_{j,t}^D \quad (2)$$

$$u_{m,t} = \tau_{ij,t}^2 \cdot u_{j,t} \quad (3)$$

Equation (1) and (2) formulate the voltage drop and branch power flow relationship, respectively. Equation (3) describes the turn ratio of OLTC. Here, we introduce new real variables $u_{i,t} = v_{i,t}^2$ to eliminate the quadratic voltage terms in equation (1) and (3). However, it is intractable to apply these primal power flow equations to OPF problems for their nonlinear and nonconvex formulations. Therefore we transform them to a three-phase LBPF model with some linearization methods.

Since the mutual impedances of branches are much smaller than the self-impedances, they can be abandoned from $\dot{z}_{ij}$ in equation (1) (2). And thus a phase-decoupled formulation (4) (5) is established, which is the three-phase distflow [19] model indeed.

$$u_{i,t}^{\varphi} - u_{m,t}^{\varphi} = 2 \cdot \left( r_{ij}^{\varphi} \cdot p_{ij,t}^{\varphi} + x_{ij}^{\varphi} \cdot q_{ij,t}^{\varphi} \right) - \left( r_{ij}^{\varphi 2} + x_{ij}^{\varphi 2} \right) \cdot \frac{p_{ij,t}^{\varphi 2} + q_{ij,t}^{\varphi 2}}{u_{i,t}^{\varphi}} \quad (4)$$

$$\begin{cases} p_{ij,t}^{\varphi} - r_{ij}^{\varphi} \cdot \dfrac{p_{ij,t}^{\varphi 2} + q_{ij,t}^{\varphi 2}}{u_{i,t}^{\varphi}} = \sum_{k \in N_d(j)} p_{jk,t}^{\varphi} + p_{j,t}^{D,\varphi} \\ q_{ij,t}^{\varphi} - x_{ij}^{\varphi} \cdot \dfrac{p_{ij,t}^{\varphi 2} + q_{ij,t}^{\varphi 2}}{u_{i,t}^{\varphi}} = \sum_{k \in N_d(j)} q_{jk,t}^{\varphi} + q_{j,t}^{D,\varphi} \end{cases} \quad (5)$$

Then, we remove the quadratic loss terms in equation (4) (5), because they are negligible compared to the branch power flow. In this manner, the LBPF model is built as follows, which is denoted as LBPF-1.

$$\textbf{LBPF-1} \begin{cases} u_{i,t}^{\varphi} - u_{m,t}^{\varphi} = 2 \cdot \left( r_{ij}^{\varphi} \cdot p_{ij,t}^{\varphi} + x_{ij}^{\varphi} \cdot q_{ij,t}^{\varphi} \right) \\ p_{ij,t}^{\varphi} = \sum_{k \in N_d(j)} p_{jk,t}^{\varphi} + p_{j,t}^{D,\varphi} \\ q_{ij,t}^{\varphi} = \sum_{k \in N_d(j)} q_{jk,t}^{\varphi} + q_{j,t}^{D,\varphi} \end{cases}$$

It should be noted that the above linearization of the primal power flow equations is realized by ignoring the mutual impedances and power losses. Therefore, the accuracy of LBPF-1 substantially relies on the network loss rate, and the errors increase as the distribution system meets heavy load or high impedance ratio conditions. LBPF-1 has been adopted in a number of ADN operation and planning problems, such as local voltage control [20] and service restoration [21], where the linearization errors are deemed to be acceptable. To further enhance the accuracy, we present another LBPF model with much higher precision in Section IV, and a three-step optimization algorithm combining these two LBPF models is proposed for implementation.

As for equation (3), the turn ratio of OLTC is formulated as equation (6), where $T_{ij,t}$ is a nonnegative integer variable representing the actual tap position of the tap-changer, $K_{ij}$ denotes the total tap number, and $\Delta \tau_{ij}$ represents the turn ratio change per tap, and $\tau_{ij}^{\min}$ is the minimal turn ratio.

$$\tau_{ij,t} = \tau_{ij}^{\min} + T_{ij,t} \cdot \Delta \tau_{ij}, 0 \leq T_{ij,t} \leq K_{ij} \quad (6)$$

The method introduced in our previous work [17] is utilized to linearize the OLTC model with exactness. The procedure is elaborated as follows.

*1)* Express the integer variable $T_{ij,t}$ in a binary expansion form (7), where $\lambda_{ij,t}^{n}$ is a binary variable and $N_{ij}$ is the length of the binary expression of $K_{ij}$.

$$T_{ij,t} = \sum_{n=0}^{N_{ij}} \left( 2^n \cdot \lambda_{ij,t}^{n} \right), \lambda_{ij,t}^{n} \in \{0,1\} \quad (7)$$

*2)* Define $\mu_{ij,t}^{\varphi} = \tau_{ij,t} \cdot u_{j,t}^{\varphi}$ and substitute a new variable $\upsilon_{ij,t}^{n,\varphi}$ for the product $\lambda_{ij,t}^{n} \cdot u_{j,t}^{\varphi}$, then have

$$\mu_{ij,t}^{\varphi} = \tau_{ij,t} \cdot u_{j,t}^{\varphi} = \tau_{ij}^{\min} \cdot u_{j,t}^{\varphi} + \sum_{n=0}^{N_{ij}} \left( 2^n \cdot \upsilon_{ij,t}^{n,\varphi} \right) \cdot \Delta \tau_{ij} \quad (8)$$

and supplement the relaxation constraint (9) to make this substitution equivalent.

$$0 \leq \upsilon_{ij,t}^{n,\varphi} \leq M_0 \cdot \lambda_{ij,t}^{n}, 0 \leq u_{j,t}^{\varphi} - \upsilon_{ij,t}^{n,\varphi} \leq M_0 \cdot \left( 1 - \lambda_{ij,t}^{n} \right) \quad (9)$$

*3)* Define $u_{m,t}^{\varphi} = \tau_{ij,t} \cdot \mu_{ij,t}^{\varphi}$ and substitute a new variable $\omega_{ij,t}^{n,\varphi}$ for the product $\lambda_{ij,t}^{n} \cdot \mu_{ij,t}^{\varphi}$, then have

$$u_{m,t}^{\varphi} = \tau_{ij,t} \cdot \mu_{ij,t}^{\varphi} = \tau_{ij}^{\min} \cdot \mu_{ij,t}^{\varphi} + \sum_{n=0}^{N_{ij}} \left( 2^n \cdot \omega_{ij,t}^{n,\varphi} \right) \cdot \Delta \tau_{ij} \quad (10)$$

and add the relaxation constraint (11) to make the substitution equivalent.

$$0 \leq \omega_{ij,t}^{n,\varphi} \leq M_0 \cdot \lambda_{ij,t}^{n}, 0 \leq \mu_{ij,t}^{\varphi} - \omega_{ij,t}^{n,\varphi} \leq M_0 \cdot \left( 1 - \lambda_{ij,t}^{n} \right) \quad (11)$$

In this manner, equation (3) is transformed equivalently into a linear formulation with equations (7)-(11).

### B. Deterministic Comprehensive Capacity Assessment Model

Taking all ANM strategies into consideration, we build a deterministic comprehensive DG capacity assessment model (DC-CAM) based on LBPF-1. It incorporates multiple time periods to deal with the daily and seasonal variability of load demands and DG outputs. Through performing the following optimization, we can obtain the maximum penetration level and optimal installation strategies of DGs, in addition to the optimal ANM schemes for each time period.

*1) Objective Function:*

$$Obj. \quad \max \sum_{i \in N_{dg}} \sum_{\varphi} y_i^{\varphi} \quad (12)$$

The objective (12) is to maximize the total DG capacity that can be integrated to ADNs.

*2) Radial Topology Constraints:*

$$s.t. \begin{cases} \sum_{i \neq j} w_{ij,t} = n_b - n_{root} \\ w_{ij,t} \in \{0,1\}, \forall t \in \Gamma \end{cases} \quad (13)$$

$$\begin{cases} \sum_{j \in N(i)} d_{ij} \cdot \left( w_{ij,t} \cdot \rho_{ij,t} \right) = -1 \\ \forall i \in N_{nr}, t \in \Gamma \end{cases} \quad (14)$$

Equation (13) ensures radial network topologies of the reconfiguration solutions. To avoid the impact of transfer buses with zero power injection, we define the nonnegative virtual power flow variable $\rho_{ij,t}$ and supplement constraint (14). $d_{ij} \in \{-1,1\}$ is the known direction parameter of power flow at branch $ij$, equaling to 1 if node $i$ is the start point while equaling to -1 if node $i$ is the end point. Indeed, equation (14) represents virtual power balance constraints at nodes, and each node has a fictitious unit load 1. To address the bilinear term in equation (14), we create a new nonnegative variable $\vartheta_{ij,t}$ to replace $w_{ij,t} \cdot \rho_{ij,t}$, and add equation (15) to make this replacement equivalent.

$$0 \leq \vartheta_{ij,t} \leq M_0 \cdot w_{ij,t}, 0 \leq \rho_{ij,t} - \vartheta_{ij,t} \leq M_0 \cdot \left( 1 - w_{ij,t} \right) \quad (15)$$

*3) DG Output Constraints:*

$$\begin{cases} p_{i,t}^{dg,\varphi} = \hat{\eta}_{i,t}^{\varphi} \cdot y_i^{\varphi} \\ \tan \theta_{i,\min}^{\varphi} \cdot p_{i,t}^{dg,\varphi} \leq q_{i,t}^{dg,\varphi} \leq \tan \theta_{i,\max}^{\varphi} \cdot p_{i,t}^{dg,\varphi} \\ \forall i \in N_{dg}, t \in \Gamma \end{cases} \quad (16)$$

DG units are assumed to operate in the maximum power point tracking mode to fully utilize renewable energy, while their reactive outputs can be adjusted continuously within the predetermined ranges.

*4) Reactive Output Constraints for VAR Compensators:*

$$\begin{cases} q_{i,\min}^{c,\varphi} \leq q_{i,t}^{c,\varphi} \leq q_{i,\max}^{c,\varphi} \\ \forall i \in N_{con}, t \in \Gamma \end{cases} \quad (17)$$

$$\begin{cases} q_{i,\min}^{d,\varphi} = \chi_{i,t}^{\varphi} \cdot \Delta q_i^d, 0 \leq \chi_{i,t}^{\varphi} \leq \chi_{i,\max}^{\varphi} \\ \forall i \in N_{dis}, t \in \Gamma \end{cases} \quad (18)$$

There are two types of VAR compensators in ADNs. Continuously adjustable VAR devices, i.e., SVCs, and discretely adjustable VAR devices, i.e., switch capacitors, are modelled as equation (17) and (18), respectively. The integer variables $\chi_{i,t}^{\varphi}$ in equation (18) can be reformulated in the following binary expansion form (19), where $M_i^{\varphi}$ is the length of the binary expression of $\chi_{i,\max}^{\varphi}$.

$$\chi_{i,t}^{\varphi} = \sum_{m=0}^{M_i^{\varphi}} \left(2^m \cdot \beta_{i,t}^{\varphi}\right), \beta_{i,t}^{\varphi} \in \{0,1\} \tag{19}$$

5) *Branch Power Flow Equation Constraints:*

$$\begin{cases} \sum_{j \in N(i)} p_{ij,t}^{\varphi} = p_{i,t}^{dg,\varphi} - \hat{p}_{i,t}^{L,\varphi} \\ \sum_{j \in N(i)} q_{ij,t}^{\varphi} = q_{i,t}^{dg,\varphi} - \hat{q}_{i,t}^{L,\varphi} + q_{i,t}^{c,\varphi} + q_{i,t}^{d,\varphi} \\ \forall (ij) \in B_{all}, t \in \Gamma \end{cases} \tag{20}$$

For the branches without a transformer:

$$\begin{cases} u_{i,t}^{\varphi} - u_{j,t}^{\varphi} \leq M_0 \cdot \left(1 - w_{ij,t}\right) + 2 \cdot \left(r_{ij}^{\varphi} \cdot p_{ij,t}^{\varphi} + x_{ij}^{\varphi} \cdot q_{ij,t}^{\varphi}\right) \\ u_{i,t}^{\varphi} - u_{j,t}^{\varphi} \geq M_0 \cdot \left(w_{ij,t} - 1\right) + 2 \cdot \left(r_{ij}^{\varphi} \cdot p_{ij,t}^{\varphi} + x_{ij}^{\varphi} \cdot q_{ij,t}^{\varphi}\right) \\ \forall (ij) \in B_{nt}, t \in \Gamma \end{cases} \tag{21}$$

For the branches with a transformer:

$$\begin{cases} (7)-(11) \\ u_{i,t}^{\varphi} - u_{m,t}^{\varphi} \leq M_0 \cdot \left(1 - w_{ij,t}\right) + 2 \cdot \left(r_{ij}^{\varphi} \cdot p_{ij,t}^{\varphi} + x_{ij}^{\varphi} \cdot q_{ij,t}^{\varphi}\right) \\ u_{i,t}^{\varphi} - u_{m,t}^{\varphi} \geq M_0 \cdot \left(w_{ij,t} - 1\right) + 2 \cdot \left(r_{ij}^{\varphi} \cdot p_{ij,t}^{\varphi} + x_{ij}^{\varphi} \cdot q_{ij,t}^{\varphi}\right) \\ \forall (ij) \in B_{tr}, t \in \Gamma \end{cases} \tag{22}$$

Equations (20)-(22) are modified from LBPF-1, and the big M method is applied to rescind the power flow constraints in disconnected branches when $w_{ij,t} = 0$.

7) *Line Thermal Capacity Constraints:*

$$\begin{cases} \left(p_{ij,t}^{\varphi}\right)^2 + \left(q_{ij,t}^{\varphi}\right)^2 \leq w_{ij,t} \cdot \left(s_{ij,\max}^{\varphi}\right)^2 \\ \forall (ij) \in B_{all}, t \in \Gamma \end{cases} \tag{23}$$

The apparent power capacity $s_{ij,\max}^{\varphi}$ of branch *ij* is used to describe the line thermal constraint, and $w_{ij,t}$ is added to limit the power flow in disconnected branches to zero. Here, several linear box constraints (24) are employed to approximate the quadratic constraint (23) with the circle constraint linearization method introduced in [22].

$$\begin{cases} -w_{ij,t} \cdot s_{ij,\max}^{\varphi} \leq p_{ij,t}^{\varphi} \leq w_{ij,t} \cdot s_{ij,\max}^{\varphi} \\ -w_{ij,t} \cdot s_{ij,\max}^{\varphi} \leq q_{ij,t}^{\varphi} \leq w_{ij,t} \cdot s_{ij,\max}^{\varphi} \\ -\sqrt{2} \cdot w_{ij,t} \cdot s_{ij,\max}^{\varphi} \leq p_{ij,t}^{\varphi} + q_{ij,t}^{\varphi} \leq \sqrt{2} \cdot w_{ij,t} \cdot s_{ij,\max}^{\varphi} \\ -\sqrt{2} \cdot w_{ij,t} \cdot s_{ij,\max}^{\varphi} \leq p_{ij,t}^{\varphi} - q_{ij,t}^{\varphi} \leq \sqrt{2} \cdot w_{ij,t} \cdot s_{ij,\max}^{\varphi} \\ \forall (ij) \in B_{all}, t \in \Gamma \end{cases} \tag{24}$$

8) *Voltage Magnitude Limit Constraints:*

$$u_{i,\min}^{\varphi} \leq u_{i,t}^{\varphi} \leq u_{i,\max}^{\varphi} \; ; \; \forall i \in N_{all}, t \in \Gamma \tag{25}$$

As a consequence, DC-CAM is formulated as a mixed integer linear programming (MILP) problem, which can be summarized as follows.

$$\textbf{DC-CAM} \begin{vmatrix} Obj. \; \max \sum_{i \in N_{dg}} \sum_{\varphi} y_i^{\varphi} \\ s.t. \; (13)-(22)(24)(25) \end{vmatrix}$$

Since DC-CAM just utilizes the predicted values of load demands $\hat{p}_{i,t}^{L,\varphi}+j\hat{q}_{i,t}^{L,\varphi}$ and DG efficiency coefficients $\hat{\eta}_{i,t}^{\varphi}$ in the assessment process, it neglects the significant uncertainty factors and needs further amelioration.

*C. Robust Comprehensive Capacity Assessment Model*

In this part, we apply the two-stage robust optimization technique to tackle the uncertainties of loads and DG outputs, and build the robust comprehensive DG capacity assessment model (RC-CAM) based on the formulation of DC-CAM. According to the profiles of historical data, we establish the polyhedral uncertainty sets (26) for uncertain DG efficiency coefficients $\tilde{\eta}_{i,t}^{\varphi}$ and load demands $\hat{p}_{i,t}^{L,\varphi}$. Here, $\Delta\underline{\eta}_{i,t}^{\varphi}$ and $\Delta\overline{\eta}_{i,t}^{\varphi}$ denote the downward and upward deviation ranges of DG efficiency coefficients respectively; $\Delta\underline{p}_{i,t}^{\varphi}$ and $\Delta\overline{p}_{i,t}^{\varphi}$ are the downward and upward deviation ranges of active load demands. Constant power factors of load demands are assumed in equation (27) to simplify analysis.

$$\Pi = \begin{cases} \tilde{\eta}_{i,t}^{\varphi} \in \left[\hat{\eta}_{i,t}^{\varphi} - \Delta\underline{\eta}_{i,t}^{\varphi}, \hat{\eta}_{i,t}^{\varphi} + \Delta\overline{\eta}_{i,t}^{\varphi}\right] \\ \tilde{p}_{i,t}^{L,\varphi} \in \left[\hat{p}_{i,t}^{L,\varphi} - \Delta\underline{p}_{i,t}^{\varphi}, \hat{p}_{i,t}^{L,\varphi} + \Delta\overline{p}_{i,t}^{\varphi}\right] \end{cases} \tag{26}$$

$$\tilde{q}_{i,t}^{L,\varphi} = \left(\hat{q}_{i,t}^{L,\varphi} / \hat{p}_{i,t}^{L,\varphi}\right) \cdot \tilde{p}_{i,t}^{L,\varphi} \tag{27}$$

The polyhedral uncertainty sets $\Pi$ (26) can be further parameterized as expression (28) with normalized deviation variables $\{\overline{a}_{i,t}^L, \underline{a}_{i,t}^L, \overline{a}_{i,t}^{dg}, \underline{a}_{i,t}^{dg}\}$, which depict the actual upward or downward deviations from the predicted values. In addition, the uncertainty budget technique [18] is utilized to control the holistic deviation level with constraint (29), which enables RC-CAM to adjust its conservativeness through tuning the budget value $B_t$. Then, $\{\hat{\eta}_{i,t}^{\varphi}, \hat{p}_{i,t}^{L,\varphi}, \hat{q}_{i,t}^{L,\varphi}\}$ in constraints (16) (20) of DC-CAM are renewed as $\{\tilde{\eta}_{i,t}^{\varphi}, \tilde{p}_{i,t}^{L,\varphi}, \tilde{q}_{i,t}^{L,\varphi}\}$.

$$\Pi' = \begin{cases} \tilde{\eta}_{i,t}^{\varphi} = \hat{\eta}_{i,t}^{\varphi} - \underline{a}_{i,t}^{dg} \cdot \Delta\underline{\eta}_{i,t}^{\varphi} + \overline{a}_{i,t}^{dg} \cdot \Delta\overline{\eta}_{i,t}^{\varphi} \\ \tilde{p}_{i,t}^{L,\varphi} = \hat{p}_{i,t}^{L,\varphi} - \underline{a}_{i,t}^{L} \cdot \Delta\underline{p}_{i,t}^{\varphi} + \overline{a}_{i,t}^{L} \cdot \Delta\overline{p}_{i,t}^{\varphi} \\ \{\overline{a}_{i,t}^L, \underline{a}_{i,t}^L, \overline{a}_{i,t}^{dg}, \underline{a}_{i,t}^{dg}\} \in [0,1] \end{cases} \tag{28}$$

$$\sum_i \left(\overline{a}_{i,t}^L + \underline{a}_{i,t}^L + \overline{a}_{i,t}^{dg} + \underline{a}_{i,t}^{dg}\right) \leq B_t \tag{29}$$

After modelling of uncertainties, we reformulate the objective function of the DG capacity evaluation issue as a two-stage robust optimization form, shown as equation (30). In the inner (second) stage, the worst-case scenarios that jeopardize DG integration are sought within the uncertainty sets, where the deviation variables $\{\overline{a}_{i,t}^L, \underline{a}_{i,t}^L, \overline{a}_{i,t}^{dg}, \underline{a}_{i,t}^{dg}\}$ are regarded as the optimization variables. Following confirmation of the worst-case scenarios, in the outer (first) stage, we make the robust optimal DG installation decisions and ANM strategies to maximize the total DG hosting capacity. In this stage, the branch status variables $w_{ij,t}$, Var injection variables $\{q_{i,t}^{dg,\varphi}, q_{i,t}^{c,\varphi}, q_{i,t}^{d,\varphi}\}$ and OLTC turn ratio variables $\tau_{ij,t}$ are also treated as the decision variables, in addition to the DG installation capacity $y_i^{\varphi}$.

$$Obj.\max_{w,q,\tau}\left[\min_a \max_y \left(\sum_{i \in N_{dg}} \sum_{\varphi} y_i^{\varphi}\right)\right] \quad (30)$$

As a consequence, our proposed RC-CAM is established as a two-stage MILP problem, shown as follows.

$$\textbf{RC-CAM} \begin{vmatrix} Obj.\max_{w,q,\tau}\left[\min_a \max_y \left(\sum_{i \in N_{dg}} \sum_{\varphi} y_i^{\varphi}\right)\right] \\ s.t. (13)\text{-}(22),(24)(25),(27)\text{-}(29) \end{vmatrix}$$

RC-CAM not only inherits the advantages of DC-CAM, which fully exploits the contributions of ANM techniques, but also can be immune to the uncertainty impacts for its robust decision-making manner. Therefore, the capacity evaluation solutions obtained via RC-CAM can guarantee optimality and robustness simultaneously, neither too conservative nor too radical.

## III. SOLUTION METHOD FOR TWO-STAGE ROBUST MODEL

In this section, the solution approach for RC-CAM is introduced. For explicit explanations, we classify all the variables into four types and use vectors to express, which involve the DG installation decision vector $y$, the power flow vector $u$, the uncertainty vector $a$ and the ANM strategy vector $z$. The detailed mapping relationship is shown as expression (31).

$$\begin{cases} \{y_i^{\varphi}\} \to y & \{w_{ij,t}, q_{i,t}^{dg,\varphi}, q_{i,t}^{con,\varphi}, q_{i,t}^{dis,\varphi}, \tau_{ij,t}\} \to z \\ \{u_{i,t}^{\varphi}, p_{ij,t}^{\varphi}, q_{ij,t}^{\varphi}, \rho_{ij,t}\} \to u & \{\overline{a}_{i,t}^L, \underline{a}_{i,t}^L, \overline{a}_{i,t}^{dg}, \underline{a}_{i,t}^{dg}\} \to a \end{cases} \quad (31)$$

Then we can express the formulation of RC-CAM in a compact form as follows.

$$Obj. \quad \max_{z \in \mathbf{Z}} \min_{a \in \Pi'} \max_{y \in \mathbf{Y}} c^\top y \quad (32)$$

$$s.t. \quad G_1 y + G_2 u + G_3 a + G_4 z + G_5 a \otimes y \leq g \quad (33)$$

$$Ia \leq b \quad (34)$$

Equation (32) and (34) are the same as objective function (30) and the uncertainty budget constraint (29), respectively, where $\mathbf{Z}$, $\Pi'$ and $\mathbf{Y}$ are the feasible sets of the associated variables. The remaining constraints in RC-CAM are expressed as equation (33). It should be mentioned that all the equality constraints are reformulated in an equivalent unified form as inequalities for notation simplicity, while $G_1$-$G_5$, $I$, $c$, $g$, and $b$ are the corresponding parameters.

Regarding the two-stage RC-CAM, we employ the wildly used column-and-constraint generation (CCG) algorithm [23] to solve this problem in iteration. According to the distinct goals in these two stages, RC-CAM is decomposed to the following master problem and subproblem, then a master-sub iterative process is conducted for solution.

### A. Master Problem of RC-CEM

$$\textbf{Master-P} \begin{vmatrix} Obj. \quad f_M = \max_{z \in \mathbf{Z}, y \in \mathbf{Y}} c^\top y \\ s.t. \begin{cases} G_1 y + G_2 u^k + G_3 a^{\circ k} + G_4 z + G_5 a^{\circ k} \otimes y \leq g \\ \forall k = 0,1,\cdots,K \end{cases} \end{vmatrix} \quad (35)$$

The master problem, shown as expression (35), is associated with the first-stage decision-making of RC-CAM. In the master problem, the uncertainty variables are realized with given $a^{\circ k}$, and finite partial enumeration scenarios with the superscript k are used to approximate the uncertainty sets $\Pi'$. Essentially, the master problem is a multi-scenario relaxation of RC-CAM, hence its optimal objective $f_M$ yields a upper bound for the original problem..

### B. Sub-Problem of RC-CEM

$$\textbf{Sub-P} \begin{vmatrix} Obj. \quad f_S = \min_{a \in \Pi'} \max_{y \in \mathbf{Y}} c^\top y \\ s.t. \begin{cases} G_1 y + G_2 u + G_3 a + G_4 z^\circ + G_5 a \otimes y \leq g \\ Ia \leq b \end{cases} \end{vmatrix} \quad (36)$$

The sub-problem, shown as expression (36), corresponds to the inner-stage optimization of RC-CAM, where a certain set of ANM scheme is confirmed with given $z^\circ$. Since the given ANM schemes may not be general optimal, the objective $f_S$ provides a lower bound for RC-CAM. The sub-problem is a minimax bi-level optimization problem, which is required to be reformulated to a monolithic form through strong duality. The dual sub-problem is derived as expression (37), where $\mu$ denotes the vector of dual variables.

$$\textbf{Dual Sub-P} \begin{vmatrix} Obj. \quad f_S = \min_{a,\mu} \mu^\top g - \mu^\top G_4 z^\circ - \mu^\top G_3 a \\ s.t. \begin{cases} G_1^\top \mu + G_5^\top \mu \otimes a = c \\ G_2^\top \mu = 0 \\ Ia \leq b, \mu \geq 0 \end{cases} \end{vmatrix} \quad (37)$$

To address the bilinear terms $\mu \cdot a$ existing in the dual sub-problem, we force the uncertainty budget $B_t$ to be an integer. In this way, the deviation variables $a$ must equal to one or zero in solutions, and the proof of this proposition is provided in [24]. Then, the aforementioned linearization method in equation (15) can be utilized to transform them into linear forms equivalently.

The master problem and the dual sub-problem are both MILP models, which can be solved efficiently by many commercially available optimizers, such as IBM CPLEX.

### C. Column-and-Constraint Generation Algorithm

Since the master problem and sub-problem provide an upper bound and a lower bound for RC-CAM respectively, it can be solved in iteration with the column-and-constraint generation algorithm, presented as follows.

1) *STEP 1* **Initialization**:

- Set $LB=0$, $UB=+\infty$; initialize the iteration counter $k=1$ and set a small tolerance $\gamma$.

- Solve DC-CAM to obtain the initial ANM strategies $z^{k-1}$ and the optimal objective $f_0$.

- Update the upper bound $UB = \min\{UB, f_0\}$.

2) **_STEP 2_ Solving Sub-problem**:

- Solve dual sub-problem with given $z^{k-1}$ to obtain the optimal $a^{\circ k}$ and objective $f_S$.

- Create new variables $u^k$ and add constraint (38) to the master problem, then update $LB = \max\{LB, f_S\}$.

$$G_1 y + G_2 u^k + G_3 a^{\circ k} + G_4 z + G_5 a^{\circ k} \otimes y \leq g \quad (38)$$

3) **_STEP 3_ Solving master problem**:

- Solve the master problem to obtain the optimal ANM strategies $z^k$ and optimal objective value $f_M$.

- Update $UB = \min\{UB, f_M\}$.

4) **_STEP 4_ Checking for convergence**:

- If $UB - LB \leq \gamma$, terminate.

- Otherwise set $k = k+1$ and go back to **_STEP 2_**.

## IV. METHOD TO IMPROVE MODEL ACCURACY

In this section, to further improve the accuracy of capacity assessment models, we present another three-phase LBPF model considering power losses, and establish an accurate DG capacity assessment model accordingly. Then, a three-step optimization algorithm is proposed to carry out the DG capacity evaluation precisely and efficiently.

### A. Three-phase LBPF Model Considering Power Losses

To obtain a three-phase LBPF model with high accuracy, we employ the approach in [25] to linearize equation (1) (2) based on the anticipative operating points $\{\dot{v}_{i,t}^0, p_{ij,t}^0, q_{ij,t}^0\}$. Firstly, the given operating voltage $\dot{v}_i^0$ is used to substitute for the voltage variables $\dot{v}_i$ in equation (1) (2). Afterwards, we approximate the quadratic loss terms around the operating power flow point $\{p_{ij,t}^0, q_{ij,t}^0\}$ with the first-order Taylor expansion, and obtain the second linear branch power flow model (LBPF-2) as follows.

$$\textbf{LBPF-2} \begin{cases} u_{i,t} - u_{m,t} = g_{ij,t}^{u,0} \cdot p_{ij,t} + b_{ij,t}^{u,0} \cdot q_{ij,t} - l_{ij,t}^{u,0} \\ g_{ij,t}^{p,0} \cdot p_{ij,t} + b_{ij,t}^{p,0} \cdot q_{ij,t} + l_{ij,t}^{p,0} = \sum_{k \in N_d(j)} p_{jk,t} + p_{j,t}^D \\ g_{ij,t}^{q,0} \cdot p_{ij,t} + b_{ij,t}^{q,0} \cdot q_{ij,t} + l_{ij,t}^{q,0} = \sum_{k \in N_d(j)} q_{jk,t} + q_{j,t}^D \end{cases}$$

Here, three-phase matrixes $g_{ij,t}^{u,0}$, $g_{ij,t}^{p,0}$, $g_{ij,t}^{q,0}$, $b_{ij,t}^{u,0}$, $b_{ij,t}^{p,0}$, $b_{ij,t}^{q,0}$ and vectors $l_{ij,t}^{p,0}$, $l_{ij,t}^{q,0}$, $l_{ij,t}^{u,0}$ are all given parameters, whose definitions and derivation process can be found in Appendix B. Power losses are taken into consideration in LBPF-2, where the linearization of primal power flow equations is realized by approximating the nonlinear terms around the operating points. Except that, no more approximations or assumptions are applied in the above process. Hence, the accuracy of LBPF-2 heavily depends on the given operating points, and it can perform very accurately and efficiently with proper operating points.

However, LBPF-2 is a state-dependent model, which cannot be applied directly to ADN assessment and planning problems, especially for network reconfiguration and OLTC regulation. Because it is difficult to obtain accurate operating points for long-term ADN optimization. Furthermore, power flow and holistic voltage profiles may change tremendously with different network topologies or secondary voltage of substations. On the contrary, LBPF-1 is a state-independent model. Hence, we take the advantages of both LBPF-1 and LBPF-2, and propose a three-step optimization algorithm to tackle this problem.

### B. Three-step Optimization Algorithm

The main idea of our proposed algorithm is utilizing the solutions of RC-CAM to provide operating points for LBPF-2. Since the errors of LBPF-1 are relatively small, which is verified by the numerical tests, RC-CAM is sufficiently accurate to serve as the generator of appropriate operating points. Besides, RC-CAM is competent to make the optimal decisions for OLTC regulation and network reconfiguration, because the discrete control variables are not sensitive to the errors of LBPF-1, and there is no need to optimize them anymore. Hence, we establish the accurate capacity assessment model (A-CAM) as follows, which evaluates the maximum DG hosting capacity with given discrete control variables and anticipative operating points.

$$\textbf{A-CAM} \begin{vmatrix} Obj. \ \max \sum_{i \in N_{dg}} \sum_{\varphi} y_i^{\varphi} \\ s.t. \begin{cases} \textbf{LBPF-2} \\ (16)-(19)(24)(25) \end{cases} \end{vmatrix}$$

The flowchart of our proposed three-step optimization algorithm is shown as Fig.2.

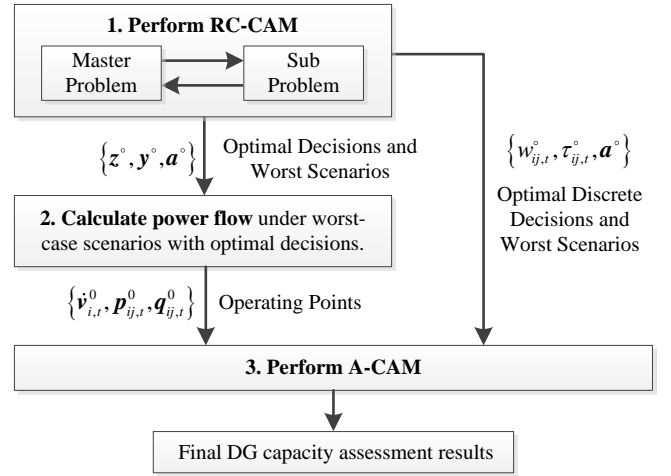

Figure 2.  The flowchart of three-step optimization algorithm.

In the first step, we perform RC-CAM to obtain the optimal ANM decisions and the worst-case scenarios. In the second step, using backward/forward sweep method or implicit Z-bus method, we calculate exact accurate power flow with the given optimal decisions. In the last step, A-CAM is implemented to procure the final DG capacity assessment results. Through this process, we can obtain much more precise assessment results for DG hosting capacity with high efficiency.

## V. NUMERICAL TESTS

In this section, we demonstrate the characteristics of RC-CAM using numerical tests on an unbalanced IEEE 33-bus distribution system. Firstly, the configurations of this test system are introduced. Secondly, we analyze the accuracy of our proposed LBPF models. Then, case studies in the worst-case scenarios are carried out to verify the robustness of RC-CAM, comparing with other two DG capacity assessment methods. Next, the optimality of RC-CAM is illustrated through Monte Carlo simulations. After that, we discuss the impact of the uncertainty budget. Lastly, the computational platform and efficiency are provided.

### A. Test System and Consideration

The modified three-phase IEEE 33-bus distribution system, shown as Fig.3, is used as our test system. There are four candidate sites (at node-10, node-17, node-22, and node-32) potentially connecting PV generators. We assume that PV-1 is three-phase integrated, PV-3 is AB-phase integrated, while PV-2 and PV-4 are just A-phase integrated; and their power factors $\cos\theta$ can be tuned from 0.78 to 0.96. The two SVCs (at node-3 and node-30) and two switch capacitors (SCs) (at node-14 and node-26) are three-phase devices. The reactive output of each SVC can be adjusted continuously from -0.1MVar to 0.15MVar for each phase. Each SC has 3 tap positions and its reactive power capacity is 0.06MVar per phase. The OLTC of the three-phase substation transformer in branch 1-2 has 5 steps enabling a ±5% turn ratio change. The five dashed red lines in Fig.3 are link lines that are usually open. To facilitate network reconfiguration, we presume that all branches in this system can be operated. The lower and upper limits of the voltage magnitude at each node are set to 0.95 p.u. and 1.05 p.u., and the apparent power capacity of each branch is 3.5 MVA per phase. More information about branch impedances and basic loads is available online [26].

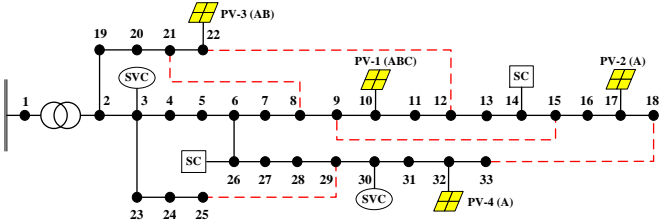

Figure 3. The modified IEEE 33-bus distribution system.

For analysis simplification, a typical day equally divided into 24 time periods serves as the multiple assessment periods. The corresponding daily PV output and load curves are shown as Fig.4. Then the real PV outputs and nodal loads are the shape coefficients of these curves multiplying the installation capacities and nominal load demands, respectively.

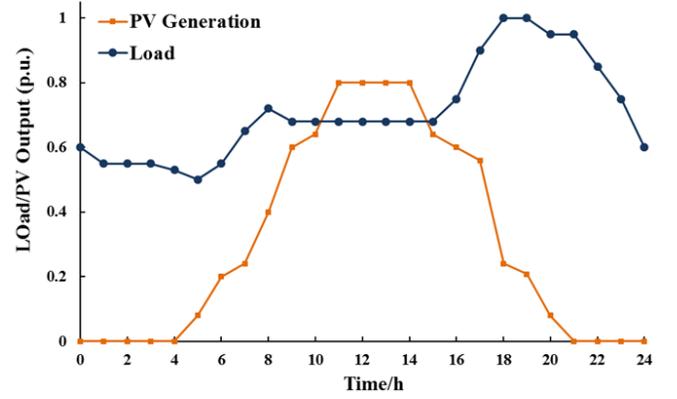

Figure 4. The daily PV output and load demand curves.

In the following tests, the predefined uncertainty set of DG efficiency coefficients and load demands were given as expression (39). We make performance comparisons among RC-CAM and two existing state-of-the-art DG capacity assessment methods [12] [15]. As mentioned above, the proposed method in [12] neglects the impacts of uncertainties, and the model in [15] does not consider the utilization of network reconfiguration. We denote the deterministic model with all ANM techniques as M-1, and denote the robust model without network reconfiguration as M-2.

$$\Pi = \begin{cases} \tilde{\eta}_{i,t}^{\varphi} \in \left[(1-\Delta_{dg})\hat{\eta}_{i,t}^{\varphi}, (1+\Delta_{dg})\hat{\eta}_{i,t}^{\varphi}\right] \\ \tilde{P}_{i,t}^{L,\varphi} \in \left[(1-\Delta_{L})\hat{P}_{i,t}^{L,\varphi}, (1+\Delta_{L})\hat{P}_{i,t}^{L,\varphi}\right] \end{cases} \quad (39)$$

### B. Accuracy Analysis of Capacity Assessment Models

Since our proposed capacity assessment models are based on LBPF-1 and LBPF-2, we applied them to calculate power flow for the test system, and analyzed their errors under different loading conditions. The backward/forward sweep method was used to provide the accurate power flow for comparison. The results of power losses and voltage errors are shown as Fig. 5 and Fig.6, respectively.

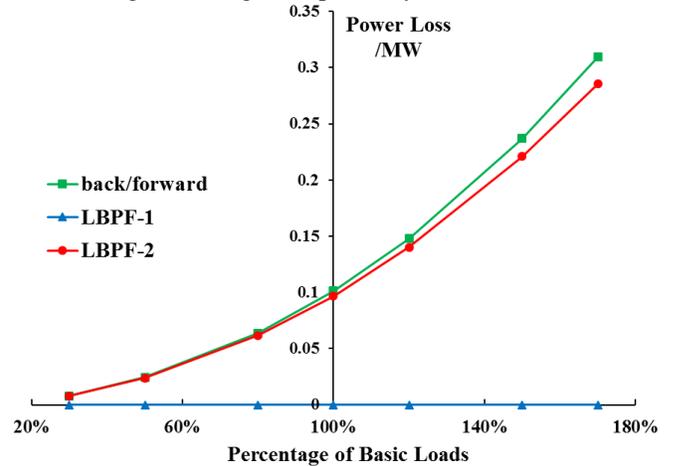

Figure 5. Power losses calculated via LBPF-1, LBPF-2 and the backward/forward sweep method under different loading conditions.

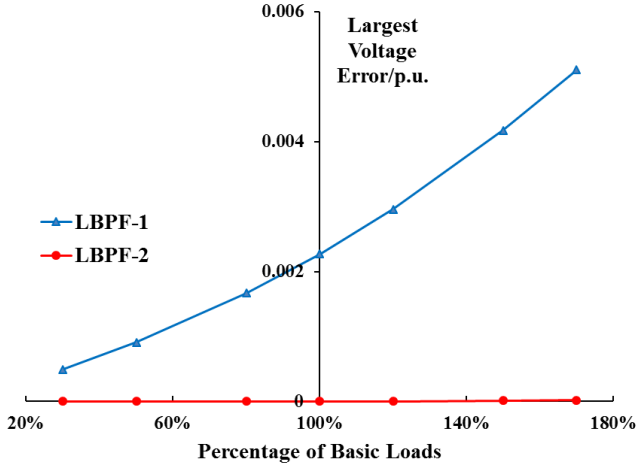

Figure 6. The largest errors of voltage magnitude with LBPF-1 and LBPF-2 under different loading conditions.

From Fig.5 and Fig.6, it is seen that LBPF-1 neglects power losses (always zero) and leads to relatively large voltage errors. While LBPF-2 can model the power losses very accurately, even in a high loading state, and the largest voltage errors of LBPF-2 are 2-3 orders of magnitude smaller than those of LBPF-1.

After that, we discussed the accuracy of our proposed DG capacity assessment models. For simplification, we assumed $\Delta_{dg} = \Delta_L = 0$, therefore RC-CAM reduced to DC-CAM. Then the three-step optimization algorithm was carried out to obtain the DG capacity assessment results via DC-CAM and A-CAM, which are presented as Table I. The true value of maximal PV hosting capacity in Table I was obtained by repeating the three-step optimization algorithm, namely using the solutions of A-CAM as new operating points for A-CAM until convergence. From Table I, it is observed that our proposed three-step optimization algorithm improved the accuracy of assessment models immensely, and A-CAM can generate very precise capacity assessment results. In the sequent tests, the three-step optimization algorithm is implemented for all assessment processes to ensure accuracy.

TABLE I
THE MAXIMAL PV HOSTING CAPACITY OBTAINED VIA DR-CAM AND A-CAM.

| Method | Maximal PV hosting capacity/MW | Absolute error/MW | Relative error |
|---|---|---|---|
| True value | 9.6795 | / | / |
| DR-CAM | 9.3789 | 0.3006 | 3.11% |
| A-CAM | 9.6814 | 0.0019 | 0.02% |

### C. Performance Comparisons in Worst-case Scenarios

In this part, we assumed $\Delta_{dg} = 0.2$, $\Delta_L = 0.15$, and removed the budget constraint (29) in RC-CAM, which implied that it sought the worst cases in the overall uncertainty set. The worst-case scenarios were extracted from the second stage of RC-CAM, and we compared the performances of RC-CAM, M-1 and M-2 in these scenarios to check their robustness. One of the time periods with highest PV efficiency (t=13h) was chosen as the test period. The hosting capacity assessment results and PV installation decisions via these three methods were presented as Table II, and the corresponding optimal schemes of network reconfiguration at t=13h were shown as Table III.

TABLE II
MAXIMAL PV HOSTING CAPACITY AND ITS ALLOCATION DECISIONS VIA THE THREE METHODS.

| Installed PV Capacity /MW | Method | nd-10 | nd-17 | nd-22 | nd-32 | Total |
|---|---|---|---|---|---|---|
| | RC-CAM | 3.399 | 0 | 3.591 | 1.211 | 8.201 |
| | M-1 | 4.082 | 0 | 4.147 | 1.452 | 9.681 |
| | M-2 | 3.978 | 0 | 3.651 | 0 | 7.629 |

TABLE III
NETWORK RECONFIGURATION SCHEMES OF THESE THREE METHODS AT T=13H.

| Method | Network Reconfiguration (Branch Opened) |
|---|---|
| RC-CAM | 28-29,32-33,8-21,9-15,12-22 |
| M-1 | 28-29,32-33,8-21,9-15,12-22 |
| M-2 | 8-21,9-15,12-22,18-33,25-29 |

Then we implemented the above PV installation decisions and ANM schemes under the worst-case scenarios. In this case, the real PV power outputs were calculated as Table IV. It is seen that although RC-CAM did not install the largest PV capacity, it achieved the highest PV outputs comparing with M-1 and M-2. Because M-1 makes decisions without consideration of uncertainties, it came up with overly radical maximum hosting capacity 9.681MW. However, the distribution system cannot absorb so much PV power injection safely, hence it led to serious voltage violation problems in the extreme conditions. As illustrated in Fig.7, when the strategies generated by M-1 were used, the voltage magnitudes of 14 nodes exceeded the upper limit, including the three nodes with PV installation. While RC-CAM and M-2 maintained voltage security due to their robust decision-making manner. While the maximum hosting capacity (7.629MW) of M-2 was too pessimistic for not utilizing network reconfiguration, therefore its real PV outputs were also lower than those of RC-CAM.

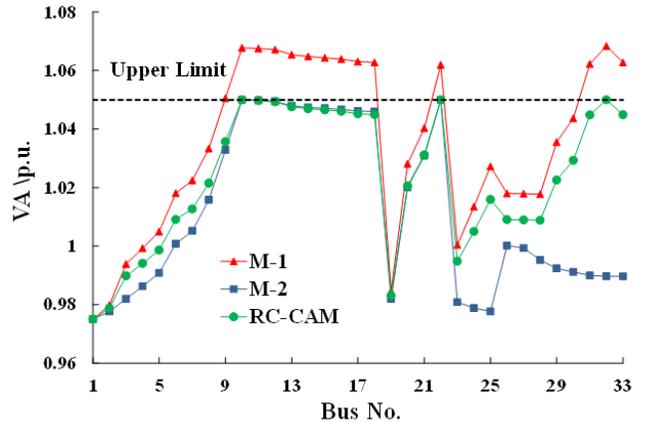

Figure 7. Voltage magnitude profiles of phase-A in the worst-case scenarios at t=13h.

TABLE IV
THE REAL PV OUTPUTS WITH THE THREE METHODS IN THE WORST-CASE SCENARIOS AT T=13H.

| Method | RC-CAM | M-1 | M-2 |
|---|---|---|---|
| Real PV Outputs/MW | 8.201 | 0 | 7.629 |

## D. Performance Comparisons via Monte Carlo Simulation

In this part, performance comparisons were made among RC-CAM, M-1 and M-2 via Monte Carlo simulation. We assumed that the actual DG outputs and load demands are independent random variables following normal distributions. Three cases with distinct predefined uncertainty sets were carried out. Up to 10000 stochastic scenarios were generated randomly and independently for each case, and $\mu \pm 3\sigma$ ($\mu$ is the expectation value and σ represents the standard deviation) intervals were ensured in sampling. The PV installation and ANM strategies obtained via these three methods were tested in these scenarios, and we calculated their average PV outputs at t=13h. Once the violation of voltage security occurred during the simulation process, we would trigger the protection mechanism and shed the connected PV units. The simulation results are summarized in Table V.

TABLE V
MONTE CARLO SIMULATION RESULTS WITH THE THREE METHODS.

| Deviation Range | Method | Total PV hosting Capacity (MW) | Average PV Outputs (MW) | Voltage Violation Rate |
|---|---|---|---|---|
| $\Delta_{dg}=0.2$ $\Delta_L=0.15$ | RC-CAM | 8.201 | 7.925 | 0 |
| | M-1 | 9.681 | 7.5633 | 39.65% |
| | M-2 | 7.629 | 7.3739 | 0 |
| $\Delta_{dg}=0.15$ $\Delta_L=0.1$ | RC-CAM | 8.546 | 8.272 | 0 |
| | M-1 | 9.681 | 7.765 | 37.43% |
| | M-2 | 7.952 | 7.699 | 0 |
| $\Delta_{dg}=0.05$ $\Delta_L=0$ | RC-CAM | 9.315 | 9.019 | 0 |
| | M-1 | 9.681 | 8.75 | 15.82% |
| | M-2 | 8.674 | 8.397 | 0 |

From Table V, it is seen that RC-CAM outperformed the other two methods with greatest average PV outputs and zero violation rate in all cases. In the terms of M-1, its PV installation decisions and ANM schemes became infeasible frequently for voltage violation, which restricted the amount of PV power outputs. The PV capacity assessment results of M-2 were overly conservative, which led to the lowest PV outputs and a waste of solar energy. While RC-CAM ensured the robustness and optimality of the capacity evaluation and ANM schemes simultaneously.

## E. Impact of Uncertainty Budget

With the uncertainty budget constraint (29), we can adjust the conservativeness of RC-CAM through tuning the value of budget. The PV capacity assessment results obtained via RC-CAM with different $B_t$ were illustrated as Fig.8. It is observed that the maximal total PV capacity has a monotonically decreasing relationship with the budget value, which is consistent with the intuitiveness that a smaller budget yields a more conservative strategy.

Besides, we can identify the choke points that limit PV connecting from the manner of uncertainty budget allocation. For instances, the budget allocation results at t=13h were presented as Table VI. When $B_{13}=1$, only $\bar{a}_{10,13}^{dg}$ was set to one after the optimization of sub-problem. It means that the upward fluctuation of PV outputs at node-10 jeopardized the maximum level of PV integration most, because the PV hosting capacity at node-10 is rather high and three-phase integrated. As $B_{13}$ became larger, the second, third etc. vital factors were picked out.

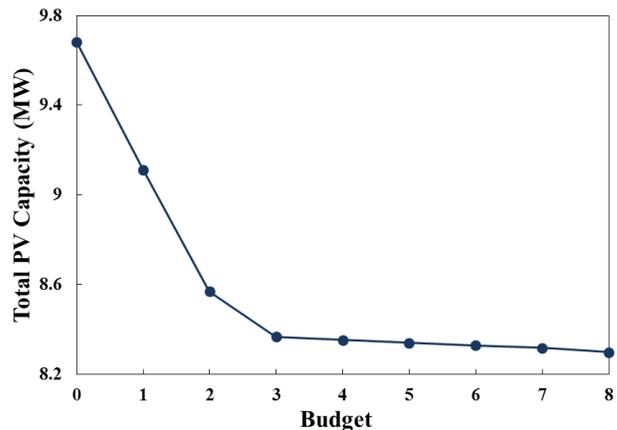

Figure 8. Total PV capacity of RC-CAM with different uncertainty budgets.

TABLE VI
UNCERTAINTY BUDGET ALLOCATION AT T=13H.

| Budget $B_{13}$ | 1 | 2 | 3 | 4 |
|---|---|---|---|---|
| Budget Allocation (others are zero) | $\bar{a}_{10,13}^{dg}=1$ | $\bar{a}_{10,13}^{dg}=\bar{a}_{22,13}^{dg}=1$ | $\bar{a}_{10,13}^{dg}=\bar{a}_{22,13}^{dg}=\bar{a}_{32,13}^{dg}=1$ | $\bar{a}_{10,13}^{dg}=\bar{a}_{22,13}^{dg}=\bar{a}_{32,13}^{dg}=\underline{a}_{14,13}^{L}=1$ |

## F. Computational Platform and Efficiency

All the numerical tests were carried out in a computational environment with Intel(R) Core(TM) i7-4510U CPUs running at 2.60 GHz with 8 GB RAM. The proposed models were programmed in Matlab 2014a and solved by an embedded IBM CPLEX 12.5 solver with the YALMIP interface. In our case studies, regarding the modified IEEE 33-bus system, it takes about 2.6 minutes to complete a RC-CAM process with the column-and-constraint generation algorithm. The CPU times for solving the master problem and sub-problem are 4.3 seconds and 51.3 seconds on average, which optimize the ANM schemes and search for the worst-case scenarios, respectively. And it usually iterates two to three times to reach convergence. Since the DG capacity assessment process is an off-line task, the computational efficiency of our proposed method is acceptable for practical applications.

## VI. CONCLUSION

To address the prediction errors of DG outputs and load demands and fully utilize the benefits of ANM techniques, we propose a robust comprehensive capacity assessment method (RC-CAM) for DGs in unbalanced distribution networks. In this method, network reconfiguration, OLTC regulation, VAR compensation and DG power factor control are all taken into consideration. Our model is formulated as a two-stage mixed integer linear programming problem after linearization. In addition, the three-step optimization algorithm is proposed to enhance the accuracy of assessment results. Numerical tests are carried out on an unbalanced IEEE 33-bus distribution system. The test results in the worst-case scenarios verify the robustness of this proposed method; and the performance comparisons against two other state-of-the-art approaches in Monte Carlo simulations demonstrate the optimality of RC-CAM. Our proposed model is a general robust comprehensive model indeed, which can be extended to many other active distribution network optimization and planning issues.

## APPENDIX

In this section, we elaborate the derivation processes of the exact three-phase branch power flow equations and LBPF-2 model. The subscript $t$ representing the time period is ignored for simplification.

### A. Derivation of Exact Three-phase Branch Power Flow

Consider branch $ij$ of radial distribution networks, shown as Fig. 1. According to Kirchhoff's current law, the power flow relationships of branch $ij$ can be formulated as

$$\dot{s}_{ij} - (\dot{v}_i - \dot{v}_m) \otimes \dot{i}_{ij}^* = \sum_{k \in N_d(j)} \dot{s}_{jk} + \dot{s}_j^D \quad (40)$$

According to Ohm's law, we have

$$\dot{v}_m = \dot{v}_i - \dot{z}_{ij} \cdot \dot{i}_{ij} = \dot{v}_i - \dot{z}_{ij} \cdot (\dot{s}_{ij}^* \oslash \dot{v}_i^*) \quad (41)$$

where $\dot{i}_{ij}$ denotes the three-phase current of branch $ij$. So we can update the power flow relationships with

$$\dot{s}_{ij} - [\dot{z}_{ij} \cdot (\dot{s}_{ij}^* \oslash \dot{v}_i^*)] \otimes (\dot{s}_{ij} \oslash \dot{v}_i) = \sum_{k \in N_d(j)} \dot{s}_{jk} + \dot{s}_j^D \quad (42)$$

Besides, we can multiply each side of equation (41) by its conjugate to eliminate the phase angles of voltage, and obtain

$$\|\dot{v}_m\|^2 = [\dot{v}_i - \dot{z}_{ij} \cdot (\dot{s}_{ij}^* \oslash \dot{v}_i^*)] \otimes [\dot{v}_i - \dot{z}_{ij} \cdot (\dot{s}_{ij}^* \oslash \dot{v}_i^*)]^*$$
$$= \|\dot{v}_i\|^2 - \dot{v}_i \otimes [\dot{z}_{ij}^* \cdot (\dot{s}_{ij} \oslash \dot{v}_i)] - [\dot{z}_{ij} \cdot (\dot{s}_{ij}^* \oslash \dot{v}_i^*)] \otimes \dot{v}_i^* \quad (43)$$
$$+ [\dot{z}_{ij} \cdot (\dot{s}_{ij}^* \oslash \dot{v}_i^*)] \otimes [\dot{z}_{ij}^* \cdot (\dot{s}_{ij} \oslash \dot{v}_i)]$$

### B. Derivation of LBPF-2 Model

As for equation (2), we define

$$\dot{z}_{ij}^\alpha = d(\dot{v}_i^{0-1}) \cdot \dot{z}_{ij} \cdot d(\dot{v}_i^{0-1})^* = r_{ij}^\alpha + jx_{ij}^\alpha \quad (44)$$

where $d(x)$ represents the diagonal matrix of vector $x$, $\dot{v}_i^{0-1} = e \oslash \dot{v}_i^0$, and we denote by $e$ the 3×1 vector of all ones.

Then the nonlinear term in equation (2) can be written as

$$[\dot{z}_{ij} \cdot (\dot{s}_{ij}^* \oslash \dot{v}_i^{0*})] \otimes (\dot{s}_{ij} \oslash \dot{v}_i^0) = d(\dot{s}_{ij}) \cdot \dot{z}_{ij}^\alpha \cdot \dot{s}_{ij}^* $$
$$= h_p(p_{ij}, q_{ij}) + jh_q(p_{ij}, q_{ij}) \quad (45)$$

where $h_p(p_{ij}, q_{ij})$ and $h_q(p_{ij}, q_{ij})$ are defined as

$$\begin{cases} h_p(p_{ij}, q_{ij}) = \\ d(p_{ij}) \cdot (r_{ij}^\alpha \cdot p_{ij} + x_{ij}^\alpha \cdot q_{ij}) + d(q_{ij}) \cdot (r_{ij}^\alpha \cdot q_{ij} - x_{ij}^\alpha \cdot p_{ij}) \\ h_p(p_{ij}, q_{ij}) = \\ d(p_{ij}) \cdot (x_{ij}^\alpha \cdot p_{ij} - r_{ij}^\alpha \cdot q_{ij}) + d(q_{ij}) \cdot (r_{ij}^\alpha \cdot p_{ij} + x_{ij}^\alpha \cdot q_{ij}) \end{cases} \quad (46)$$

We linearize equation (46) around the operating point $\{p_{ij,t}^0, q_{ij,t}^0\}$ with the first-order approximation, and thus obtain

$$\begin{cases} g_{ij}^{p,0} \cdot p_{ij} + b_{ij}^{p,0} \cdot q_{ij} + l_{ij}^{p,0} = \sum_{k \in N_d(j)} p_{jk} + p_j^D \\ g_{ij,t}^{q,0} \cdot p_{ij} + b_{ij}^{q,0} \cdot q_{ij} + l_{ij}^{q,0} = \sum_{k \in N_d(j)} q_{jk} + q_j^D \end{cases} \quad (47)$$

where

$$\begin{cases} g_{ij}^{p,0} = e - d(p_{ij}^0) \cdot r_{ij}^\alpha - d(r_{ij}^\alpha \cdot p_{ij}^0) - d(x_{ij}^\alpha \cdot q_{ij}^0) + d(q_{ij}^0) \cdot x_{ij}^\alpha \\ b_{ij}^{p,0} = e - d(p_{ij}^0) \cdot x_{ij}^\alpha - d(r_{ij}^\alpha \cdot q_{ij}^0) + d(x_{ij}^\alpha \cdot p_{ij}^0) - d(q_{ij}^0) \cdot r_{ij}^\alpha \\ l_{ij}^{p,0} = d(r_{ij}^\alpha \cdot p_{ij}^0) \cdot p_{ij}^0 + d(x_{ij}^\alpha \cdot q_{ij}^0) \cdot p_{ij}^0 \\ \qquad + d(r_{ij}^\alpha \cdot q_{ij}^0) \cdot q_{ij}^0 - d(x_{ij}^\alpha \cdot p_{ij}^0) \cdot q_{ij}^0 \\ g_{ij,t}^{q,0} = e - d(p_{ij}^0) \cdot x_{ij}^\alpha - d(x_{ij}^\alpha \cdot p_{ij}^0) + d(r_{ij}^\alpha \cdot q_{ij}^0) - d(q_{ij}^0) \cdot r_{ij}^\alpha \\ b_{ij}^{q,0} = e - d(q_{ij}^0) \cdot x_{ij}^\alpha - d(r_{ij}^\alpha \cdot p_{ij}^0) - d(x_{ij}^\alpha \cdot q_{ij}^0) + d(p_{ij}^0) \cdot r_{ij}^\alpha \\ l_{ij}^{q,0} = d(x_{ij}^\alpha \cdot p_{ij}^0) \cdot p_{ij}^0 - d(r_{ij}^\alpha \cdot q_{ij}^0) \cdot p_{ij}^0 \\ \qquad + d(r_{ij}^\alpha \cdot p_{ij}^0) \cdot q_{ij}^0 + d(x_{ij}^\alpha \cdot q_{ij}^0) \cdot q_{ij}^0 \end{cases} \quad (48)$$

As for equation (1), we define

$$\dot{z}_{ij}^\beta = (\dot{v}_i^0 \cdot \dot{v}_i^{0-T})^* \otimes \dot{z}_{ij} = r_{ij}^\beta + jx_{ij}^\beta \quad (49)$$

$$\dot{z}_{ij}^\gamma = \dot{z}_{ij} \cdot d(\dot{v}_i^{0-1})^* = r_{ij}^\gamma + jx_{ij}^\gamma \quad (50)$$

Then, the nonlinear terms in equation (1) can be written as

$$\dot{v}_{i,t}^0 \otimes [\dot{z}_{ij}^* \cdot (\dot{s}_{ij,t} \oslash \dot{v}_{i,t}^0)] + [\dot{z}_{ij} \cdot (\dot{s}_{ij,t}^* \oslash \dot{v}_{i,t}^{0*})] \otimes \dot{v}_{i,t}^{0*}$$
$$= 2 \cdot (r_{ij}^\beta \cdot p_{ij} + x_{ij}^\beta \cdot q_{ij}) \quad (51)$$

$$h_u(p_{ij}, q_{ij}) = [\dot{z}_{ij} \cdot (\dot{s}_{ij,t}^* \oslash \dot{v}_{i,t}^{0*})] \otimes [\dot{z}_{ij}^* \cdot (\dot{s}_{ij,t} \oslash \dot{v}_{i,t}^0)]$$
$$= (r_{ij}^\gamma \cdot p_{ij}) \otimes (r_{ij}^\gamma \cdot p_{ij}) + (x_{ij}^\gamma \cdot q_{ij}) \otimes (x_{ij}^\gamma \cdot q_{ij})$$
$$+ (r_{ij}^\gamma \cdot q_{ij}) \otimes (r_{ij}^\gamma \cdot q_{ij}) + (x_{ij}^\gamma \cdot p_{ij}) \otimes (x_{ij}^\gamma \cdot p_{ij}) \quad (52)$$
$$+ 2 \cdot (r_{ij}^\gamma \cdot p_{ij}) \otimes (x_{ij}^\gamma \cdot q_{ij}) - 2 \cdot (r_{ij}^\gamma \cdot q_{ij}) \otimes (x_{ij}^\gamma \cdot p_{ij})$$

After that, $h_u(p_{ij}, q_{ij})$ is linearized around the operating point $\{p_{ij,t}^0, q_{ij,t}^0\}$ with the first-order Taylor expansion, and we obtain

$$u_i - u_m = g_{ij}^{u,0} \cdot p_{ij} + b_{ij}^{u,0} \cdot q_{ij} - l_{ij}^{u,0} \quad (53)$$

where

$$\begin{cases} g_{ij}^{u,0} = 2 \cdot r_{ij}^\beta - f_p^u(p_{ij}^0, q_{ij}^0) \\ b_{ij}^{u,0} = 2 \cdot x_{ij}^\beta - f_q^u(p_{ij}^0, q_{ij}^0) \\ l_{ij}^{u,0} = h_u(p_{ij}^0, q_{ij}^0) - f_p^u(p_{ij}^0, q_{ij}^0) \cdot p_{ij}^0 - f_q^u(p_{ij}^0, q_{ij}^0) \cdot q_{ij}^0 \end{cases} \quad (54)$$

and

$$\begin{cases} f_p^u(p_{ij}^0, q_{ij}^0) = 2 \cdot \begin{bmatrix} d(x_{ij}^\gamma \cdot q_{ij}^0) \cdot r_{ij}^\gamma - d(r_{ij}^\gamma \cdot q_{ij}^0) \cdot x_{ij}^\gamma \\ + d(r_{ij}^\gamma \cdot p_{ij}^0) \cdot r_{ij}^\gamma + d(x_{ij}^\gamma \cdot p_{ij}^0) \cdot x_{ij}^\gamma \end{bmatrix} \\ f_q^u(p_{ij}^0, q_{ij}^0) = 2 \cdot \begin{bmatrix} d(r_{ij}^\gamma \cdot p_{ij}^0) \cdot x_{ij}^\gamma - d(x_{ij}^\gamma \cdot p_{ij}^0) \cdot r_{ij}^\gamma \\ + d(x_{ij}^\gamma \cdot q_{ij}^0) \cdot x_{ij}^\gamma + d(r_{ij}^\gamma \cdot q_{ij}^0) \cdot r_{ij}^\gamma \end{bmatrix} \end{cases} \quad (55)$$

We refer the reader to Reference [25] for more information about the above deviation process.